# Critical couplings in topological-insulator waveguide-resonator systems observed in elastic waves

Si-Yuan Yu[†,1,2*], Cheng He[†,1,2], Xiao-Chen Sun[†,1], Hong-Fei Wang[1], Ji-Qian Wang[1], Zi-Dong Zhang[1], Bi-Ye Xie[1], Yuan Tian[1], Ming-Hui Lu[1,2*], and Yan-Feng Chen[1,2*]

1. *National Laboratory of Solid State Microstructures & Department of Materials Science and Engineering, Nanjing University, Nanjing, Jiangsu 210093, China*
2. *Jiangsu Key Laboratory of Artificial Functional Materials, Nanjing University, Nanjing 210093, China*

*Correspondence to: yusiyuan@nju.edu.cn ; luminghui@nju.edu.cn ; yfchen@nju.edu.cn

[†]These authors contributed equally to this work.

**ABSTRACT**

Waveguides and resonators are core components in the large-scale integration of electronics, photonics, and phononics, both in existing and future scenarios. In certain situations, there is critical coupling of the two components; *i.e.*, no energy passes through the waveguide after the incoming wave couples into the resonator. The transmission spectral characteristics resulting from this phenomenon are highly advantageous for signal filtering, switching, multiplexing, and sensing. In the present study, adopting an elastic-wave platform, we introduce topological insulator (TI), a remarkable achievement in condensed matter physics over the past decade, into a classical waveguide-ring-resonator configuration. Along with basic similarities with classical systems, a TI system has important differences and advantages, mostly owing to the spin-momentum locked transmission states at the TI boundaries. As an example, a two-port TI waveguide resonator can fundamentally eliminate upstream reflections while completely retaining useful transmission spectral characteristics, and maximize the energy in the resonator, with possible applications being novel signal processing, gyro/sensing, lasering, energy harvesting, and intense wave–matter interactions, using phonons, photons, or even electrons. The present work further enhances the confidence of using topological protection for practical device performance and functionalities, especially considering the crucial advantage of introducing (pseudo)spins to existing conventional configurations. More in-depth research on advancing phononics/photonics, especially on-chip, is foreseen.

**INTRODUCTION**

Topological insulators (TIs) of classical waves (*e.g.*, in the fields of electromagnetics and mechanics) stem from exciting developments in the field of solid-state electronic materials[1,2], which are insulating in their interior but can support conducting states on their surface. By constructing pseudospins and corresponding spin-momentum locking in a bosonic system, analogous quantum spin Hall effects of, for example, photons and phonons, have been successfully realized in recent years[3,4]. A series of milestone works has offered revolutionarily transmission channels with backscattering immunity (*i.e.*, robustness to defects such as fabrication imperfections and arbitrary bends) on a variety of platforms such as those for optics[5], microwaves[6–8], plasmonics[9–11], silicon photonics[12–14], mechanical waves[15], airborne sound[16-18], and elastic waves[19-22]. All of these channels are passive and free of an external field. Following the wide invention of these ideal transmission channels, a thought-provoking application-driven question is whether spectral functions can be implemented inside the channels. Specifically, it is asked whether there is a TI resonator solution that matches these revolutionary TI channels. Inspired by the vigorous development of ring resonators in, for example, photonics[23–25], a potentially effective approach is to wrap the channels themselves into closed loops, building similar TI "ring" resonators. Referring to conventional waveguide-ring-resonators, mostly in optics, spectral features resulting from the iconic critical coupling phenomenon[26–28] (*i.e.*, no energy passing through the waveguide after the incoming wave is coupled into the ring-resonator, with the typical transmission spectrum $S_{21}$ showing a near-to-zero dip at the resonance frequency) serve as basic functions in advanced signal filtering, switching, multiplexing, and high-sensitivity gyro/sensing. Recently, TI "ring" resonance has been demonstrated for topological nonlinear harmonic output and on-chip topological lasers[29–32]. However, the spectral characteristics of a TI waveguide-resonator system or even a single "ring" resonator, no matter the physical platform, remain elusive.

Meanwhile, in conventional waveguide-resonator systems, the occurrence of critical couplings always increases reflectance in the input channel owing to inevitable backscattering in practice[28,33–36], which further induces both intra- and interchannel crosstalk (noise) in an integrated system with multiple connections[37–39], largely reducing performance. It has been found that this kind of noise is dominant in many functional scenarios; *e.g.*, scenarios of representative resonator gyros[33]. As the level of integration increases, more functional units (*e.g.*, resonators, splitters, and multiplexers) are connected in a single chip; hence, the reflection/crosstalk accumulates and may even result in rapid failure of system functions. Unlike the electronic system, a passive integrated photodiode (or phonon diode) has not been put into practical use thus far, although many notable attempts have been made[40–42]. Therefore, avoiding input reflections and induced crosstalk (noise), especially in spectral functional devices, poses a challenge for further development of large-scale integrated photonic and phononic circuitry.

In the present paper, using an elastic wave system, we systematically analyse a "ring" resonator constructed using a closed TI boundary, and demonstrate experimentally TI waveguide-resonator systems with advantageous spectral characteristics. We find that, unlike conventional ring resonators, a single TI "ring" resonator supports both travelling-wave whispering-gallery modes (WGMs)[26,28,43,44] and unavoidable split standing wave modes (SWMs)[45–48] simultaneously at separate frequencies. These two types of mode support different spin quantum numbers (±½ and 0, respectively) and need to meet different conditions for critical coupling to the TI waveguide, allowing the possibility of using them separately with different functionalities. In a scenario in which a TI-WGM resonator is coupled to a TI waveguide, because both support the same spins ±½ locked with momentum, if the initial state of the system has only one spin, then back reflections with the opposite spin are unable to be excited, even in the case of critical coupling; *i.e.*, $S_{11}$ is always zero even when $S_{21}$ is close to zero. Conversely, in a scenario in which a TI-SWM cavity is coupled to a TI waveguide, because the spin-less SWM (spin 0) can be converted with each of the spin ±½ modes, even if the initial state of the system has only one arbitrary spin, eventually, all three spins (+½, 0, and −½) including the backscattering can be excited. There is always input reflection when critical coupling occurs (*i.e.*, non-zero peaks appear in $S_{11}$ when $S_{21}$ drops close to zero), similar to conventional scenarios. These similarities and unique advantages that we have found with conventional waveguide-resonator systems lay the foundation for future technologies using artificial TIs in both photonics and phononics. Meanwhile, basic but unprecedented solid-state acoustic devices for future integrated phononic circuitry (*e.g.*, add-drop filters) are presented.

## RESULTS AND DISCUSSION

### *TI "ring" resonators (cavities) and their eigenstates*

Our experimental and numerical investigation begins with an independent TI "ring" resonator. With the same implementation as in our previous study[19], we used a perforated plate to construct both a TI and an ordinary insulator (OI) of mechanical elastic waves. The TI and OI have the same lattice constant ($a = 3a_0$) of hexagonal unit cells, and each cell contains six of the same perforated holes in a plain plate with the same thickness. The only difference is the hole–centre distance, $b$, which equals $a_0$ (OI, **Supplementary Fig. 4a**) and $1.12a_0$ (TI, **Supplementary Fig. 4b**). An elastic topologically protected boundary is constructed using an adjacent TI–OI interface (**Fig. 1a**). Two gapless helical (*i.e.*, spin-momentum locked) edge states appear in the bulk band gap (**Fig. 1b**), accompanied by elastic (pseudo)spin +½ and −½.

If this TI–OI boundary is bent and constitutes a closed loop (**Fig. 1c**), then it can be considered a TI "ring" resonator, or for short, a TI cavity. When a one-dimensional (1-D) boundary evolves into a zero-dimensional (0-D) loop, its energy band also decreases from 1-D to 0-D, accompanied by an eigenstate decrease from a frequency-continuous band to discrete points (**Fig. 1d**)[49–51]. Note that the occurrence and number of eigenstates is highly correlated with the cavity size; particularly, a larger cavity results in a greater number of supported eigenstates[49] (see **Supplementary Fig. 5**). These discrete eigenstates are generally symmetrically distributed above and below the central frequency of the helical edge states ($f_c$) and always appear in pairs with the same spatial symmetry[50]. On the basis of their spatial symmetry (see **Supplementary Fig. 6**), we label these paired eigenstates as modes "…, 3 (3′), 2 (2′), 1 (1′), 0 (0′), …". They can be generally divided into two categories, and those in the same category are symmetric about $f_c$.

In one category, eigenstates in pairs have the same frequency; *e.g.*, modes 1 (1') and 2 (2') in this paper. These degenerate eigenstates are actually the WGMs in the TI system. Each pair can be interpreted as the superposition of one TI waveguide mode with spin +½ (circulating clockwise at the cavity boundary) and one with spin −½ (circulating counter clockwise), with the same amplitude but different phases. When the TI cavity operates at these degenerate frequencies, the cavity can be simply thought of as a WGM resonator, supporting a travelling wave mode(s) with spin +½ (and/or spin −½) circulating around the cavity boundary.

In the other category, the eigenstates in pairs no longer have the same frequency; *e.g.*, modes 0 (0´) and 3 (3´). Obviously, they can no longer be regarded as a superposition of two independent circulating modes with spin +½ and spin −½ around the cavity boundary. Their appearance is essentially due to the inevitable reduced geometric symmetry of the closed TI cavity boundary loop (see the theoretical deduction in **Supplementary Note I**), also observed in non-TI systems[52,53], and the final form of their presentation is indeed spilt SWMs[45–48], which have been systematically studied in optical microcavities, especially for sensing. When the TI cavity operates at these nondegenerate frequencies, the cavity should be treated as a whole instead of a closed loop, supporting only the collective resonance of the standing wave throughout its region. The degree of frequency splitting of these SWMs varies depending on the material systems and structures. In the elastic case of the present study, the degree of mode 0 (0') splitting ($\Delta f/f_c = 0.16\%$) is more pronounced than that of mode 3 (3') ($\Delta f/f_c = 0.06\%$), as shown in **Fig. 1e** and **Fig. 1f**. As the cavity size increases, the frequency splitting gradually decreases (see **Supplementary Fig. 5**).

Note that for conventional ring resonators, eigenstates can be all WGMs but no SWMs. In the TI cavity, however, the appearance of SWMs is unavoidable, simply because the spatial symmetry of any TI cavity is always broken. Taking the TI cavity in this paper as an example, because the unit cell of the phononic crystal is hexagonal, even when the TI cavity is to be designed extremely large and to contain a great many cells, there is still no way to make it perfectly circular. Therefore, if the cavity geometry is irregular, the appearance of SWMs will be confusing and difficult to predict. To control (or reduce) the appearance of these unavoidable SWMs, a reasonable approach is to design the cavity to have a certain high symmetry. As an example, the cavity in this paper is designed to be hexagonal, consistent with its own unit cell. It can also be designed as a diamond or triangle. The appearance of SWMs can still be predicted through similar symmetry analysis as in **Supplementary Note I**.

### *Mode conversions in the TI waveguide resonator and possible critical couplings*

To demonstrate the importance of TI design for a waveguide-resonator system, especially the pseudospins ±½ and the two categories of TI cavity modes as just mentioned, we compared mode conversion in a conventional waveguide-resonator system and that in a TI waveguide-resonator system.

**Fig. 2a** shows the mode conversions in a conventional waveguide-resonator system. Normally, the resonator works in the most common WGMs. In the waveguide, the conversion of same-frequency forward (+$k$) and backward (−$k$) propagating modes is possible owing to backscattering. Similarly, in the resonator, the conversion of clockwise and counterclockwise travelling (*i.e.*, circulating) WGMs is possible. When the waveguide and resonator are close and coupled, the two opposite propagating modes in the waveguide can convert to the two opposite WGMs in the resonator. In such a two-port waveguide-resonator system, only one initial mode is needed to excite all other modes at the same frequency finally. As an example, as shown in **Fig. 2b**, when there is only one left incidence into the two-port system, part of the incident wave flows into the cavity; waves in the cavity can also flow back to the waveguide through two paths: 1) a forward path, exiting the system from the right port, and 2) a backward path, via lumped backscattering, exiting the system from the left port. In most cases, this system transmittance is close to 1 and the reflectance is close to zero owing to presently available high-quality cavities with tiny backscattering. Only when critical coupling occurs, owing to Fabry–Pérot interference, is the forward path prohibited and the transmittance drops to zero. The backward path alone, in contrast, becomes important, with backscattering inevitably being enhanced, leading to a reflectance that is no longer zero (as in **Fig. 2c**) [28,33-36].

In comparison, **Fig. 2d** shows mode conversions in a TI waveguide-resonator system. In the TI waveguide, the same frequency forward (+$k$, with a pseudo spin +½) and backward (−$k$, with a pseudo spin −½) propagating modes cannot convert to each other owing to spin-momentum locking at the TI boundary.

WGMs in the TI cavity are travelling waves and their behaviours are thus almost identical with the those of TI waveguide modes but have a closed loop. Additionally, clockwise and counter-clockwise travelling (*i.e.*, circulating) WGMs cannot convert to each other, and their spin numbers remain ±½. Here, when the waveguide and resonator are coupled, mode conversions between them can only occur between the modes with the same pseudo spins; *i.e.*, from spin +½ (−½) to spin +½ (−½) and vice versa. This results in the entire system being without a spin flipping mechanism. If we initially excite a specific spin mode in the system, whether it is a forward propagating mode in the TI waveguide or a WGM in the TI cavity, the final system only has modes with the same spin. For example, as shown in **Fig. 2e**, when there is still only one left incidence into the two-port TI system, part of the incident wave still flows into the cavity. However, waves in the cavity only flow back to the waveguide via the forward path alone and finally exit the system from the right port alone. Imaging that under such a scenario, if critical coupling (*i.e.*, Fabry–Pérot interference) occurs, the incident waves/energy in the cavity have not even one channel through which to flow out, resulting in advantageous spectral characteristics of both zero transmittance and zero reflectance at the same frequencies (as in **Fig. 2f**).

In the case of SWMs, because 1) an SWM has only one for each resonance frequency and 2) the backscattering suppression in the TI waveguide, linear mode conversions only occur between the TI waveguide and TI cavity. In contrast with TI-WGMs, a TI-SWM can be excited by both the forward (+$k$) and backward (−$k$) propagating modes of the waveguide, simply because of the standing-wave nature. With existing knowledge, an SWM is spin-less (spin 0) and can be regarded as zero-dimensional collective resonance on the whole. Naturally, the excited SWM can be further converted into two opposite waveguide modes at the same time. Therefore, in this case, although backscattering between the two opposite travelling modes is suppressed in the TI waveguide, the conversion from spin +½ to spin −½ and vice versa can be achieved through the TI-SWM serving as a bridge. For example, as shown in **Fig. 2g**, excited waves in the cavity (spin 0) from the left incidence (spin +½) can flow back to the waveguide through both the right (spin +½) and left (spin −½) paths. Imagining that critical coupling (*i.e.*, Fabry–Pérot interference) occurs in this scenario, the spectral characteristics will be similar to those of the conventional case, except for split frequencies (as in **Fig. 2c**).

*Observation of critical couplings in the TI waveguide-resonator system*

To investigate and verify the two categories of TI cavity modes (*i.e.*, TI-WGMs and TI-SWMs), we conducted a comparative experiment, separately exhibiting their critical couplings under two different configurations. A two-port TI waveguide-resonator system was built. The left side of a sample was a vertical straight TI waveguide while the right side was a hexagonal TI cavity, as shown in **Figs. 3a** and **4a**. The entire system had only two ports, one at the bottom (P1) and the other at the top (P2). We placed a longitudinal wave transducer on the bottom port to excite elastic waves of different frequencies to travel up the straight waveguide. As the elastic wave passed through the adjacent region of the TI waveguide and TI cavity, a portion of the energy entered the cavity and possibly resonated there. The energy in the cavity may also have flowed out, returning to the straight waveguide and eventually flowing out of the system from the upper port. We measured the elastic energy spectrum inside the cavity to determine if any cavity mode was excited. The transmittance of the two-port system ($S_{21}$) was measured to determine whether critical coupling occurred. Experiments and numerical simulations showed that, for the two categories of TI cavity mode, the spatial relationships (in particular, the spacings) between the cavity and waveguide required for critical coupling are different.

In the TI-SWM case, critical coupling occurs when the cavity is relatively close to the waveguide. An example is shown **Fig. 3a**, in which the spacing between the two components is only two layers of molecules (considering a six-hole unit cell as a molecule). All major calculated cavity modes are excited in the experiment (**Fig. 3b**), although the splitting of modes 3 and 3' is somewhat elusive (which may be because the degree of mode splitting is further reduced when the cavity and waveguide are coupled at a close distance). Modes 1 and 2 (*i.e.*, TI-WGMs) are excited but do not critically couple with the waveguide, while mode 3 (3') and mode 0 (0') (*i.e.*, TI-SWMs) have critical couplings with the straight waveguide (**Fig. 3c**), accompanied by increased input reflection (**Fig. 3d**), similar to the case for all conventional waveguide resonator systems. The measured energy field distributions (**Fig. 3e** to **3h**) further confirm the conclusion drawn for spectral measurements; *i.e.*, for modes 3 (3') and 0 (0'), although the energy flowing into the cavity does not flow out of the system at the upper port when critical coupling occurs, it is not completely lost in the cavity but returns to the original path and exits the system at the bottom port, confirming the occurrence of spin flipping from elastic (pseudo)spin +½ to spin −½ via the spin-0 TI-SWM of the cavity.

In the TI-WGM case, by contrast, critical coupling occurs when the TI cavity is relatively far from the waveguide, as shown in **Fig. 4a**, with a spacing of eight layers of molecules. Again, all major calculated cavity modes are excited (**Fig. 4b**). This time, however, modes 0 (0') and 3 (3') (*i.e.*, TI-SWMs) do not critically couple with the waveguide, while modes 1 and 2 (*i.e.*, TI-WGMs) have critical couplings with the straight waveguide (**Fig. 4c**). Unprecedentedly, even if critical coupling occurs, no obvious input reflections are detected (**Fig. 4d**). Measurements of the energy field distribution (**Fig. 4e** to **4h**) further demonstrate this peculiar phenomenon; *i.e.*, the energy flowing into the cavity does not flow out to exit the system from the upper port, nor does it return to the original incident path, and it is instead all stored (and dissipated) in the cavity, causing the energy density inside the cavity to sharply increase. The cavity serves as a perfect energy absorber and looks like an acoustic "black hole".

Except for negligible air damping, the energy loss of the mechanical system (especially that of the cavity) is mainly 1) intrinsic material loss and 2) radiation loss outward from the sample boundary (absorbed by the sound absorbing material around the sample). Limited by the overall size of our experimental samples, the distance from the cavity boundary to the sample boundary is still not large enough, and considerable radiation loss thus occurs in the present system. In the case that the sample boundary is far enough away from the cavity (*e.g.*, in the future, the whole system is expected to be integrated into a microchip, with a much wider TI (or OI) area allowed outside the cavity), because the energy is exponentially attenuated inside the OI (or TI), the external radiation loss can be theoretically reduced to zero. That is to say, all energy input into the system will be dissipated in only the form of intrinsic material loss. This scenario is advantageous for all conventional waveguide-resonator configurations; *e.g.*, in photonic integrated circuits where waveguides and cavities are coupled using evanescent waves, through which there are always radiation losses. Hence, this new mechanism indicates the possibility of improving the Q value of the coupled waveguide resonator. Although the Q values measured in the present paper are close to $10^4$ at frequencies of approximately 110 kHz, this is most likely due to 1) the high thermoelastic dissipation (one kind of intrinsic material loss) of the metal sample material and 2) large radiation loss in our experiment. In future integrated on-chip scenarios, materials with lower intrinsic loss, such as Si, $Si_3N_4$, AlN, and $LiNbO_3$, can be fully adopted and are expected to bring about much higher *f*·Q values. At the same time, the higher energy capacity/density of the cavity will facilitate the use of this device in fields such as energy harvesting, sensing, phonon lasing, and wave–matter interaction enhancement, using solid-state acoustic waves.

The fact that TI-SWM resonators require smaller spacing than TI-WGM resonators to realize critical couplings with the TI waveguide can be understood in three respects. 1) The criterion for the critical coupling based on classical coupled-mode theory is that the waveguide-resonator coupling loss equals the intrinsic resonator loss[27]. 2) The former loss is related to the spacing of the waveguide resonator; *i.e.*, smaller spacing results in stronger coupling and greater coupling loss. 3) The latter loss relates to different TI cavity modes. TI-WGM resonators and corresponding TI waveguide-resonator systems are backscattering suppressed; TI-SWM resonators, in contrast, bring backscattering into the waveguide resonator system, such that energy is more easily consumed and there is greater equivalent intrinsic resonator loss than for TI-WGM resonators.

*Four-port add-drop filter with zero input reflection*

Another example of further using such a TI coupled waveguide resonator is a four-port add-drop filter. We experimentally built such a prototype (**Fig. 5a**). The left and right sides of the sample are two parallel straight TI–OI waveguides with a TI cavity sandwiched between them. The overall system has four ports, one each at the bottom left (P1), upper left (P2), bottom right (P3) and upper right (P4). The spacing between the TI cavity and two waveguides is chosen to be eight layers; *i.e.*, the spacing of the demonstrated critically coupled TI-WGMs in the previous two-port system.

This time, we placed the transducer at the bottom left (P1) port to excite elastic waves to travel up the left TI waveguide and then measured the energy spectrum in the cavity, transmittance at the three output ports ($S_{21}$, $S_{31}$ and $S_{41}$) and reflectance at the incident port ($S_{11}$). Unsurprisingly, for this configuration, although all cavity modes are excited (**Fig. 5b**), only the TI-WGMs (*i.e.*, modes 2 and 1) are critically coupled to the left straight waveguide. When critical couplings occur, the two-port transmittance of the left waveguide ($S_{21}$), serving as a band-stop filter, falls to nearly zero, and this is still accompanied by zero input reflection (**Fig. 5c**). At the same time, all incoming energy is transferred, via the cavity, to the bottom-right port (P3), serving as a band-pass filter, with hardly any energy transfer to the upper-right port (P4) of the right waveguide (**Fig. 5d**). The measured energy field distributions (**Fig. 5e** and **5f**) well confirm the results of the spectral measurements.

In fact, the above four ports have fully equivalent positions; *i.e.*, results are similar regardless of which port energy is incident on. As an example, when energy is incident on the upper-right port (P4), it completely transfers to the upper-left port (P2) via TI-WGM cavity resonance. With this device, one can easily add (or remove, as a consequence of reciprocity) a signal to a bus waveguide (**Fig. 5g**); *i.e.*, implement an add-drop filter that serves as an important element in multiplexers, modulators and switches[54–56]. However, in this case, there is zero input reflection and induced crosstalk owing to the TI design.

**CONCLUSION AND OUTLOOK**

We first systematically analysed a ring cavity constructed using a TI boundary and studied its coupling with the TI waveguide. While having many similarities with conventional waveguide resonator systems, TI systems have important differences and advantages. We proposed and experimentally confirmed two categories of TI cavity modes that simultaneously exist in a single cavity by observing their independent critical couplings with the TI waveguide under different spacings. TI-SWMs serve as spin flippers, converting upward spin +½ to downward spin −½. TI-WGMs are of particular advantage when their critical couplings occur: 1) reflections are completely eliminated while the required transmission spectral characteristics are retained and 2) the incident energy is bound entirely inside the cavity with no channel through which to exit, resulting in an extremely high energy capacity (density). Both of these advantages have been highly anticipated in conventional waveguide resonator systems.

Although our experiments are based on a system of mechanical waves, the findings are also fully applicable to electromagnetic waves, especially in integrated (topological) photonics[13,14,29–32,57]. At the same time, as a solid-state phononic system, our experimental platform can be fully scaled down, adopting piezoelectric[58–63], non-piezoelectric[20,22,64–66], or hybrid[67] materials, to work in or above the GHz band and paving the way for integrated phononic circuitry[20,67,68] with various functionalities (*e.g.*, phonon lasering[69,70], sensing and signal processing[71]) in both classical and quantum[63,66,72,73] regions.

A challenging shortcoming is the limitation of the relatively small bulk band gap width of the existing TI (4.5% in this work and 13% as the currently reported maximum[20]), although we can obtain multiple TI-WGMs/SWMs in the band gap by increasing the size of the TI cavity (see **Supplementary Fig. 5**) and sacrificing the free spectral range (an important indicator in integrated optics) of the cavity modes. If the working bandwidth of these TI systems can be effectively increased, it will greatly facilitate our demonstrating functionalities in practical scenarios. Of course, the introduction of anomalous spin-momentum locking phenomenon may not necessarily have to rely on TIs[74–79]. This work may open an avenue for introducing (pseudo)spins into traditional transmission devices, especially in integrated systems, hopefully stimulating more in-depth research with a similar concept.

# METHODS

## Sample preparation

Samples in this work were prepared exclusively on polished stainless-steel plates (type 201, mass density of ~7900 kg/m$^3$) with a fixed plate thickness of ~7.80 mm. Their elastic parameters were determined adopting the ultrasonic scattering echo method; *i.e.*, a Young's modulus ~205 GPa and Poisson's ratio of ~0.32. The plates were perforated on a precision computer-numerical-controlled milling machine to create phononic crystals with an identical hole radius $r = 3.52$ mm. An impedance-matched soundproof adhesive, comprising epoxy resin, graphite, and tungsten powder, was coated on the circumferences of the samples (black regions shown in Figs. 3a, 4a, and 5a) to prevent unwanted solid-state acoustic reflections.

## Numerical calculation

Calculations of the mechanical/elastic band structure and eigen mode distributions shown in this work were conducted adopting a three-dimensional finite element method and the structural mechanics module of the commercial software COMSOL MULTIPHYSICS.

## Experimental apparatus

Broadband piezoelectric ultrasonic transducers (having a central frequency of 110 kHz) were attached to the sample surface as excitation sources. They were driven by a function generator followed by a power amplifier. A fiber-laser-based two-wave mixing interferometer with a lock-in amplifier was used to interrogate amplitude/phase information of the back-reflected optical measurement beam from the sample surface. The field profile of the mechanical wave was then mapped out in a point-by-point fashion according to the information of the reflected optical measurement beam.

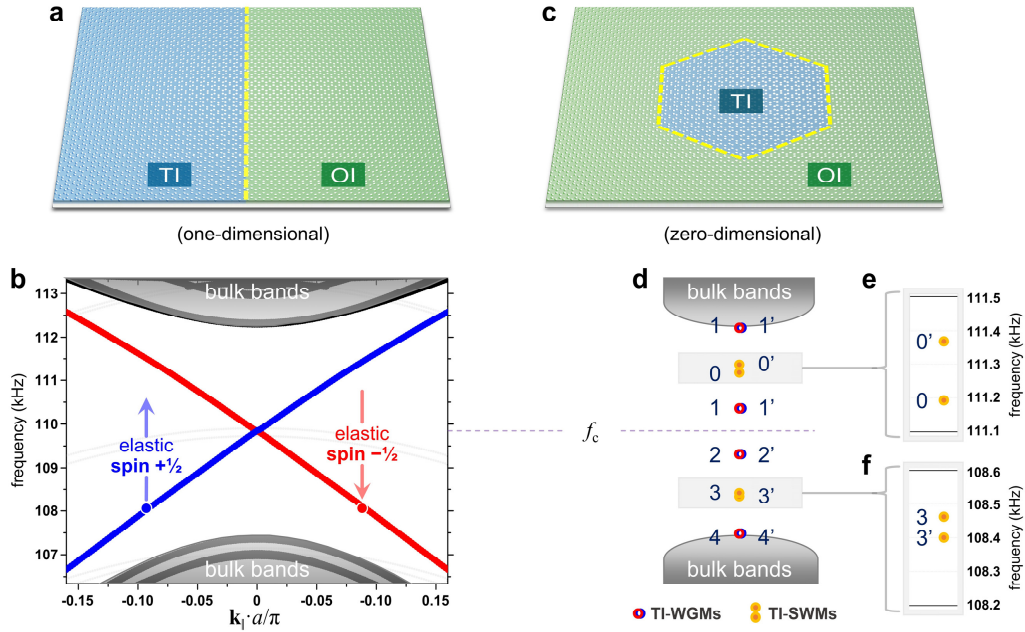

**Figure 1 | TI "ring" resonator (*i.e.*, TI cavity) and its paired eigenstates of two categories**

**a**: Schematic of an elastic topologically protected boundary comprising an elastic TI adjacent to an elastic OI. **b**: Calculated elastic band structure of the 1-D TI–OI boundary. Two gapless helical (*i.e.*, spin-momentum locked) edge states appear in the bulk band gap. **c**: Schematic of an elastic TI cavity comprising a hexagonal TI embedded in an OI. **d**: Calculated elastic band structure of the hexagonal TI cavity. When the 1-D TI–OI boundary evolves into a closed loop, its energy band changes from 1-D to 0-D. The eigenstates change from a frequency-continuous band to discrete points, paired and symmetrically distributed above and below the central frequency of the helical edge states ($f_c$), labelled as "…, 3 (3'), 2 (2'), 1 (1'), 0 (0'), …". The eigenstates can be divided into two categories, with those in the same category being symmetric about $f_c$, according to whether they are degenerate. Degenerate paired eigenstates are TI-WGMs (*e.g.*, 1 (1') and 2 (2')) while frequency-split paired eigenstates are TI-SWMs (*e.g.*, 1 (1') and 2 (2')), respectively shown in **e** and **f**.

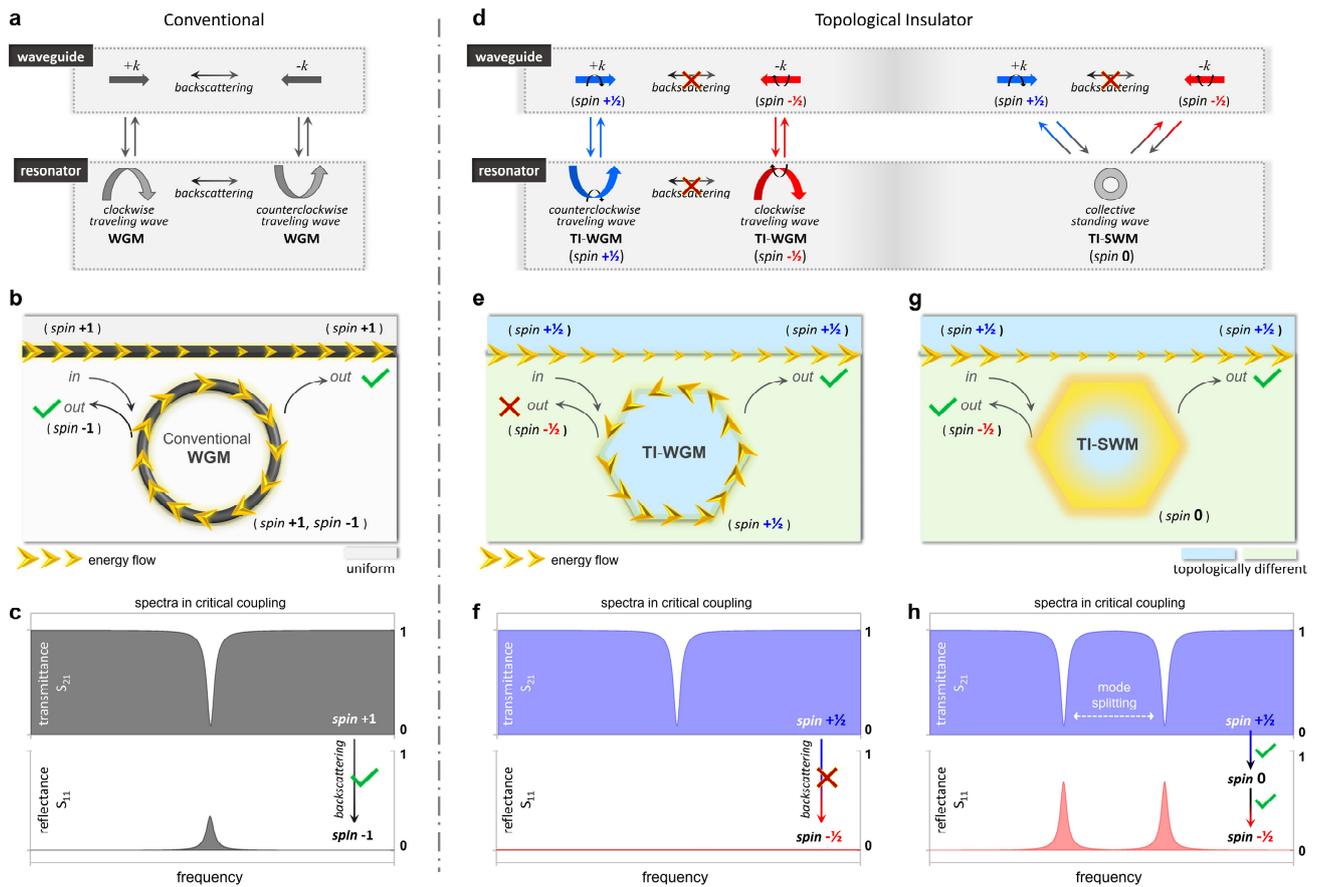

**Figure 2 | Conventional waveguide resonator vs. TI waveguide resonator**

**a**: Mode conversions, as the arrows indicate, in a conventional waveguide resonator. Only one initial mode is needed to excite all other modes at the same frequency finally. **b**: When there is only one remaining incidence into this two-port system, part of the incident wave flows into the cavity. Waves in the cavity flow back to the waveguide through via two paths: 1) the forward path, exiting the system from the right port, and 2) the backward path, via lumped backscattering, exiting the system from the left port. **c**: When there is critical coupling, the forward path is prohibited owing to Fabry–Pérot interference, and the transmittance thus drops to zero. The backward path alone, in contrast, becomes important, with backscattering inevitably being enhanced, leading to reflectance. **d:** Mode conversions, as the arrows indicate, in a TI waveguide resonator. **e**: In the case of TI-WGMs, waves still flow into the cavity when there is a left incidence. However, waves in the cavity flow back to the waveguide via the forward path alone and finally exit the system from the right port alone. **f**: If there is critical coupling, incident waves in the cavity have not even one channel through which to flow out, potentially resulting in simultaneous zero transmittance and reflectance at the same frequency. **g**: In the case of TI-SWMs, excited standing waves in the cavity from the left incidence (spin +½) flow back to the waveguide through both right (spin +½) and left (spin −½) paths. **h**: If there is critical coupling, the spectral characteristics will be similar to those in the conventional case, except for the split frequencies.

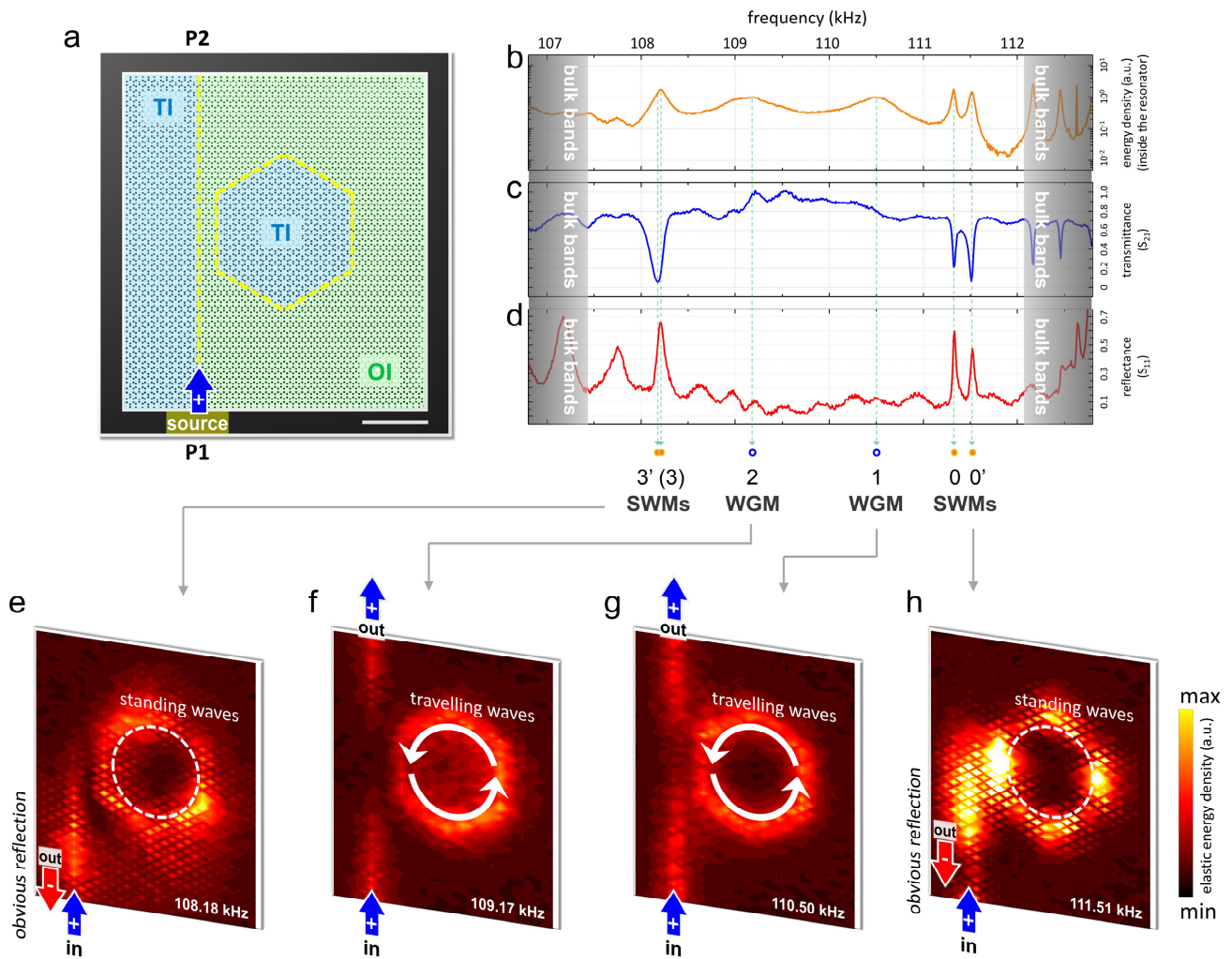

**Figure 3 | Selective critical couplings of TI-SWMs to the TI waveguide, serving as spin flippers**

**a**: Sample image and experimental configuration. The left side of the sample is a vertical straight TI waveguide, with a port at the bottom (P1) and a port at the top (P2), a hexagonal TI cavity on the right side and a spacing of two molecular layers. Elastic waves of different frequencies, covering the entire TI/OI bulk band gap, are excited at port P1 to travel up the straight waveguide, with a portion of the waves entering the cavity, resonating or flowing back. **b**: Measured energy spectrum inside the cavity, confirming that all major calculated cavity modes are experimentally excited, although the splitting of modes 3(3') is somewhat elusive. **c** and **d**: Measured spectra of the transmission coefficient ($S_{21}$) and reflection coefficient ($S_{11}$) of the two-port system reinforced by measured energy field distributions under the excitation of modes (**e**) 3 (3'), (**f**) 2, (**g**) 1, and (**h**) 0 (0'). Modes 1 and 2, although excited, cannot be critically coupled to the waveguide, as evidenced by the still-considerable system transmittance at resonant frequencies. Mode 3 (3') and mode 0 (0') can be critically coupled to the waveguide modes, as evidenced by the distinct drops in $S_{21}$ as well as the increase in $S_{11}$ at resonant frequencies and also evidenced by the measured energy field distributions. The TI cavity now becomes a spin flipper, converting upward spin +½ to downward spin −½. In this scenario, the measured Q factor is about 800 for mode 3/3' and about 2900 for mode 0/0'.

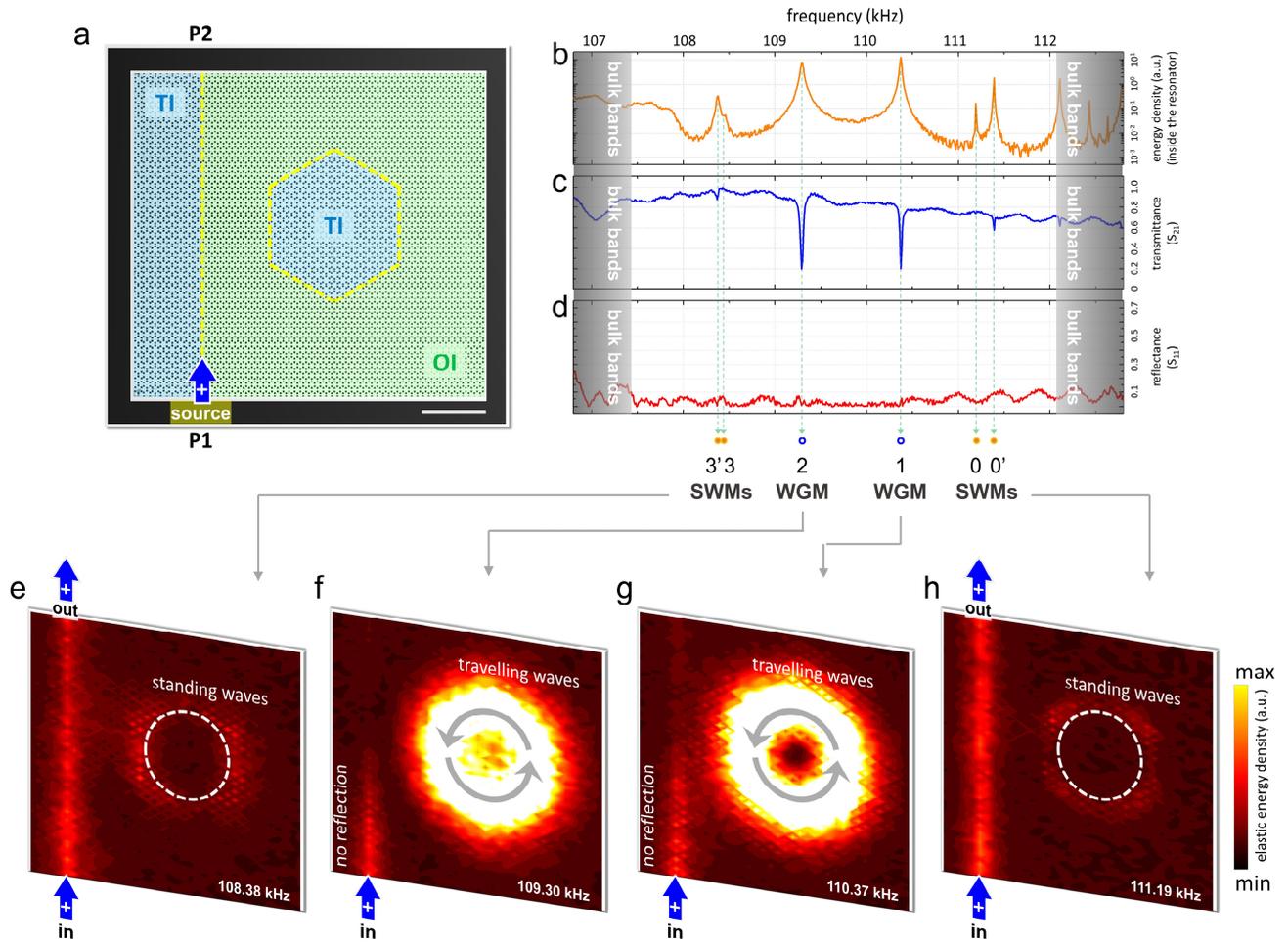

**Figure 4 | Selective critical couplings of TI-WGMs to the TI waveguide, serving as perfect absorbers**

**a**: Sample image and experimental configuration. The left side of the sample is a vertical straight TI waveguide, with a port at the bottom (P1) and a port at the top (P2), a hexagonal TI cavity on the right side and a spacing of eight molecular layers. Elastic waves of different frequencies, covering the entire TI/OI bulk band gap, are excited at port P1 to travel up the straight waveguide, with a portion of the waves entering the cavity, resonating or flowing back. **b**: Measured energy spectrum inside the cavity, confirming that all major calculated cavity modes are excited. Compared with the results shown in Fig. 3, owing to the different spacing between the waveguide and cavity, the frequencies and intensities of the excited modes are slightly changed. **c** and **d**: Measured spectra of the transmission coefficient ($S_{21}$) and reflection coefficient ($S_{11}$) of the two-port system reinforced by measured energy field distributions under the excitation of modes (**e**) 3 (3'), (**f**) 2, (**g**) 1, and (**h**) 0 (0'). Modes 3 (3') and 0 (0'), although excited, are relatively faint and cannot be critically coupled to the waveguide, as evidenced

<br />

by the still-considerable system transmittance at resonant frequencies. Modes 2 and 1 can be critically coupled to the waveguide modes, as evidenced by the distinct drops in $S_{21}$ at resonant frequencies, with no apparent increase in reflectance, and also evidenced by the measured energy field distributions. The entire TI cavity, amazingly, now becomes a perfect energy absorber with greatly enhanced energy density. In this scenario, the measured Q factor is about 3100 for mode 2/2' and about 6100 for mode 0/0'.

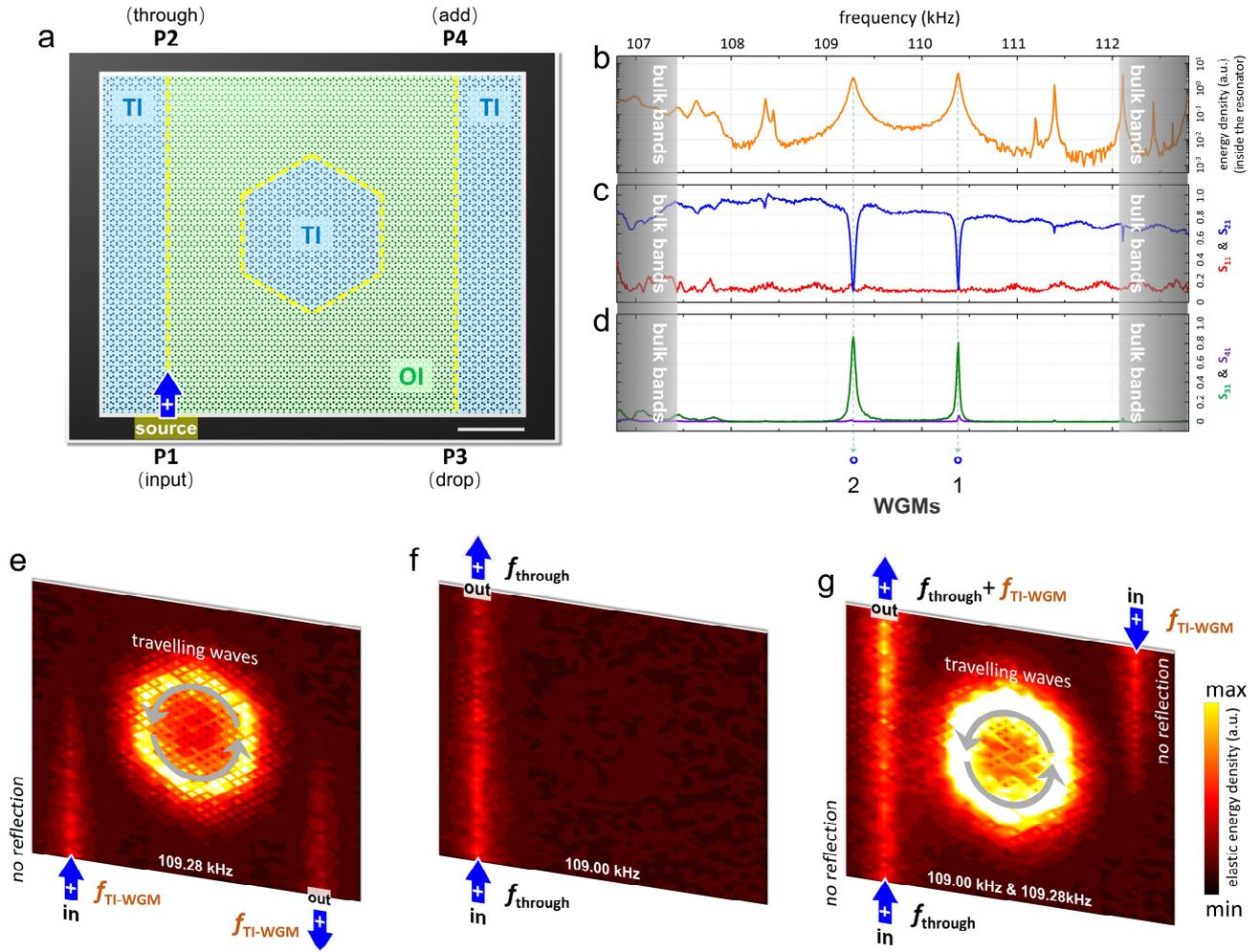

**Figure 5 | TI four-port add-drop filter**

**a**: Sample image and experimental configuration. Two TI waveguides are placed in parallel on the left and right sides of the sample, leading to four ports in the system: bottom left (P1), upper left (P2), bottom right (P3) and upper right (P4). The hexagonal TI cavity serves as a bridge, placed between the two parallel waveguides with the same spacing as that in Fig. 4 for eight molecular layers. Elastic waves of different frequencies, covering the entire TI/OI bulk band gap, are excited at port P1 to initially travel up the left straight waveguide, with a portion of the waves entering the cavity, resonating or flowing elsewhere. **b**: Measured energy spectrum inside the cavity, confirming that all major cavity modes are experimentally excited, with almost the same frequencies as in Fig. 4. **c** and **d**: Measured spectra of the transmission and reflection coefficients of the four-port system, $S_{11}$, $S_{21}$, $S_{31}$ and $S_{41}$, reinforced by measured energy field distributions at (**e**) excitation frequencies of the TI-WGMs (*e.g.*, at $f_{\text{TI-WGM}}$ = 109.28 kHz) and (**f**) at any arbitrary frequency other than those of these TI-WGMs (*e.g.*, $f_{through}$ = 109.00 kHz). **h**: Measured energy field distributions when elastic waves of $f_{through}$ are incident at P1 and those of $f_{TI-WGM}$ are incident at P4. The two different incident waves completely converge when exiting the system from P2.

# Supplementary Information

*for*

# Critical couplings in topological insulator waveguide-resonator systems observed in elastic waves

Si-Yuan Yu[†1,2*], Cheng He[†1,2], Xiao-Chen Sun[†1], Ji-Qian Wang[1], Zi-Dong Zhang[1], Bi-Ye Xie[1], Hong-Fei Wang[1], Yuan Tian[1], Ming-Hui Lu[1,2*], and Yan-Feng Chen[1,2*]

[1]*National Laboratory of Solid State Microstructures & Department of Materials Science and Engineering, Nanjing University, Nanjing, Jiangsu 210093, China.*

[2]*Jiangsu Key Laboratory of Artificial Functional Materials, Nanjing University, Nanjing 210093, China*

[†]*These authors contributed equally to this work.*

*\*e-mail: yusiyuan@nju.edu.cn; luminghui@nju.edu.cn; yfchen@nju.edu.cn.*

---

This Supplementary Information includes:

➢ **Supplementary Note I**: Microscopic Theory for the TI Cavity

——Part I: Paired Degenerate Eigenstates

　　（Supplementary Figs. 1 and 2）

——Part II: Paired Eigenstates with Broken Degeneracy

　　（Supplementary Fig. 3 & Tables S1, S2, and S3）

➢ **Supplementary Note II**: Other Supplements for the Experiments

——Supplementary Fig. 4 | Structures of elastic OI and TI used in this research

——Supplementary Fig. 5 | TI cavities of different sizes and their eigenmodes

——Supplementary Fig. 6 | Paired cavity modes shown in Fig. 1 of the main text

——Supplementary Fig. 7 | Cavity mode excitations of two different categories

# Supplementary Note I: Microscopic Theory for the TI Cavity

## ——Part I: Paired Degenerate Eigenstates

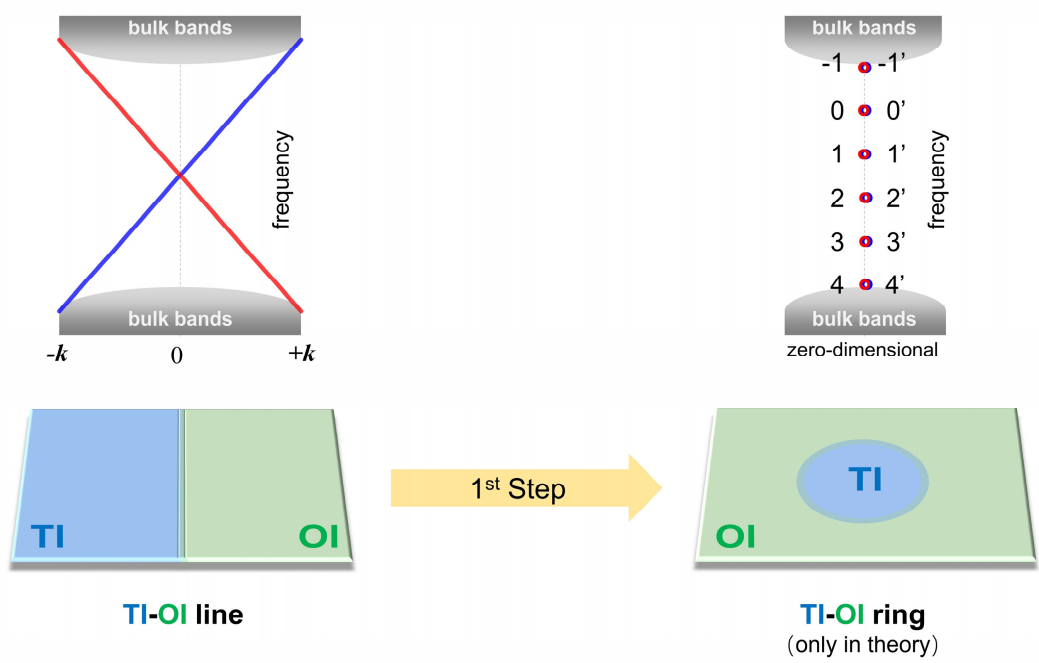

**Supplementary Fig. 1 |** From 2D helical edge states to paired degenerate eigenstates

In our implementation, if we choose the basis $(p_+, d_+, p_-, d_-)$, then the effective $k \cdot p$ Hamiltonian is expressed as

$$H = \begin{bmatrix} M + Bk^2 & Ak_+ & & \\ A^* k_- & -M - Bk^2 & & \\ & & M + Bk^2 & Ak_- \\ & & A^* k_+ & -M - Bk^2 \end{bmatrix}, \quad (S1)$$

where $k_\pm = k_x \pm i k_y$ and $k^2 = k_x^2 + k_y^2$. $A = -i|A|$ is a purely imaginary number. $B$ and $M$ are real numbers. We next consider a finite-sized sample with circular geometry. Under polar coordinates, we have

$$\begin{aligned} k_x &= \cos\phi \, k_r - \sin\phi \, k_\phi \\ k_y &= \sin\phi \, k_r + \cos\phi \, k_\phi \end{aligned}, \quad (S2)$$

in which

$$k_x = -i\partial_x, \, k_y = -i\partial_y, \, k_r = -i\partial_r, \, k_\phi = (1/r)(-i\partial_\phi), \quad (S3)$$

such that

$$k^2 = k_x^2 + k_y^2 = -(\partial_x^2 + \partial_y^2) = -\left(\partial_r^2 + \frac{1}{r^2}\partial_\phi^2 + \frac{1}{r}\partial_r\right), \quad (S4)$$

$$k_{\pm} = k_x \pm i k_y = \left(\cos\phi k_r - \sin\phi k_\phi\right) \pm i\left(\sin\phi k_r + \cos\phi k_\phi\right) = e^{\pm i\phi} k_r \pm i e^{\pm i\phi} k_\phi. \tag{S5}$$

Our discussion focuses on the interface, leading to $r \to R$, where $R$ is the position of the interface, as shown in **Supplementary Fig. 2**. Noting that the wave decays from the interface, we assume that $k_r = -i\partial_r \to -i\kappa, \kappa \in \mathbb{R}$ and $\kappa \gg 1/R$, where $\kappa$ is the decay rate of the field.

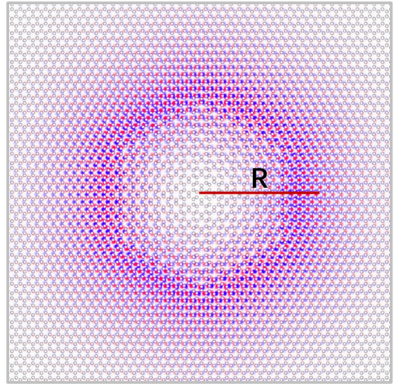

**Supplementary Fig. 2** | Field distribution of a representative topological cavity eigenstate.

We therefore ignore the term $\dfrac{1}{R^2}$ and assume

$$k^2 = -\left(\partial_r^2 + \frac{1}{r^2}\partial_\phi^2 + \frac{1}{r}\partial_r\right) \to -\left(\kappa^2 + \frac{1}{R}\kappa + \frac{1}{R^2}\partial_\phi^2\right) \approx -\left(\kappa^2 + \frac{1}{R}\kappa\right) \to -\left(\partial_r^2 + \frac{1}{R}\partial_r\right). \tag{S6}$$

Substituting Eqs. S2, S4, and S5 into Eq. S1 and separating the radial component of k from the transverse component, we have

$$H = \begin{bmatrix} M & Ae^{i\phi}k_r & & \\ A^*e^{-i\phi}k_r & -M & & \\ & & M & Ae^{-i\phi}k_r \\ & & A^*e^{i\phi}k_r & -M \end{bmatrix} + \begin{bmatrix} -B\left(\partial_r^2 + \dfrac{1}{R}\partial_r\right) & iAe^{i\phi}k_\phi & & \\ -iA^*e^{-i\phi}k_\phi & B\left(\partial_r^2 + \dfrac{1}{R}\partial_r\right) & & \\ & & -B\left(\partial_r^2 + \dfrac{1}{R}\partial_r\right) & -iAe^{-i\phi}k_\phi \\ & & iA^*e^{i\phi}k_\phi & B\left(\partial_r^2 + \dfrac{1}{R}\partial_r\right) \end{bmatrix}$$

$$. \tag{S7}$$

The second part of Eq. S5 can be regarded as perturbations, and we neglect the small second-order amount. We have

$$H_0 = \begin{bmatrix} M & Ae^{i\phi}k_r & & \\ A^*e^{-i\phi}k_r & -M & & \\ & & M & Ae^{-i\phi}k_r \\ & & A^*e^{i\phi}k_r & -M \end{bmatrix}, \tag{S8}$$

$$\Delta H = \begin{bmatrix} -B\left(\partial_r^2 + \dfrac{1}{R}\partial_r\right) & iAe^{i\phi}k_\phi & & \\ -iA^*e^{-i\phi}k_\phi & B\left(\partial_r^2 + \dfrac{1}{R}\partial_r\right) & & \\ & & -B\left(\partial_r^2 + \dfrac{1}{R}\partial_r\right) & -iAe^{-i\phi}k_\phi \\ & & iA^*e^{i\phi}k_\phi & B\left(\partial_r^2 + \dfrac{1}{R}\partial_r\right) \end{bmatrix}. \tag{S9}$$

With the basis $(p_+, d_+, p_-, d_-)$ transformed to $(e^{i\phi}p_+, d_+, e^{-i\phi}p_-, d_-)$, the unperturbed Hamiltonian is

$$H_0 = \begin{bmatrix} M & Ak_r & & \\ A^*k_r & -M & & \\ & & M & Ak_r \\ & & A^*k_r & -M \end{bmatrix}. \tag{S10}$$

In fact, because of the different topologies, the mass terms are different inside $(M(r<R) = +M_1 > 0)$ and outside $(M(r>R) = -M_2 < 0)$ the interface. Using the conditions $\psi(r \to R^-) = \psi(r \to R^+)$ and $\psi(r \to 0) = \psi(r \to \infty) = 0$, we obtain the eigenvalue and eigenwave functions:

$$\omega_0 = 0 \tag{S11}$$

$$|1\rangle = \begin{pmatrix} e^{i\phi}p_+ & d_+ & e^{-i\phi}p_- & d_- \end{pmatrix} \begin{pmatrix} 1 \\ 1 \\ 0 \\ 0 \end{pmatrix} F(r) = \begin{pmatrix} p_+ & d_+ & p_- & d_- \end{pmatrix} \begin{pmatrix} e^{i\phi} \\ 1 \\ 0 \\ 0 \end{pmatrix} F(r), \tag{S12}$$

$$|2\rangle = \begin{pmatrix} e^{i\phi}p_+ & d_+ & e^{-i\phi}p_- & d_- \end{pmatrix} \begin{pmatrix} 0 \\ 0 \\ 1 \\ 1 \end{pmatrix} F(r) = \begin{pmatrix} p_+ & d_+ & p_- & d_- \end{pmatrix} \begin{pmatrix} 0 \\ 0 \\ e^{-i\phi} \\ 1 \end{pmatrix} F(r), \tag{S13}$$

with

$$F(r) = \sqrt{\dfrac{M_1 M_2}{|A|(M_1 + M_2)}} \begin{cases} e^{\frac{M_1}{|A|}(r-R)}, & r < R \\ e^{-\frac{M_2}{|A|}(r-R)}, & r > R \end{cases}, \tag{S14}$$

$$\int_0^\infty F(r)^2 \, dr = 1/2. \tag{S15}$$

Considering the perturbations,

$$H = \begin{pmatrix} \langle 1|\Delta H|1\rangle & \langle 1|\Delta H|2\rangle \\ \langle 2|\Delta H|1\rangle & \langle 2|\Delta H|2\rangle \end{pmatrix} = \begin{pmatrix} |A|k_\phi + \dfrac{|A|}{2R} & 0 \\ 0 & |A|k_\phi - \dfrac{|A|}{2R} \end{pmatrix} \to \begin{pmatrix} |A|\dfrac{m}{R} + \dfrac{|A|}{2R} & 0 \\ 0 & -|A|\dfrac{m}{R} + \dfrac{|A|}{2R} \end{pmatrix}. \quad (S16)$$

We here quantize the wavevector $k_\phi$:

$$k_\phi = \frac{-i\partial_\phi}{R} \to \frac{m}{R} \quad (S17)$$

and

$$\langle 1|\Delta H|1\rangle = \int_0^\infty dr \begin{pmatrix} e^{-i\phi} & 1 & 0 & 0 \end{pmatrix} F(r) \begin{bmatrix} B\left(\partial_r^2 + \dfrac{1}{R}\partial_r\right) & iAe^{i\phi}k_\phi & & \\ -iA^* e^{-i\phi}k_\phi & -B\left(\partial_r^2 + \dfrac{1}{R}\partial_r\right) & & \\ & & B\left(\partial_r^2 + \dfrac{1}{R}\partial_r\right) & -iAe^{-i\phi}k_\phi \\ & & iA^* e^{i\phi}k_\phi & -B\left(\partial_r^2 + \dfrac{1}{R}\partial_r\right) \end{bmatrix} \begin{pmatrix} e^{i\phi} \\ 1 \\ 0 \\ 0 \end{pmatrix} F(r)$$

$$= \int_0^\infty dr\left(F(r) B\left(\partial_r^2 + \dfrac{1}{R}\partial_r\right)F(r) - F(r) B\left(\partial_r^2 + \dfrac{1}{R}\partial_r\right)F(r)\right) + \left(\int_0^\infty F(r)^2 dr\right)\left(e^{-i\phi}(iAe^{i\phi}k_\phi) + (-iA^*e^{-i\phi}k_\phi)(e^{i\phi})\right)$$

$$= \frac{1}{2}\left(iAk_\phi - iA^*k_\phi - i\dfrac{A^*}{R}\right) = |A|k_\phi + \dfrac{|A|}{2R}$$

$$. \quad (S18)$$

We finally obtain the eigenvalues

$$\boxed{\omega = \omega_0 + \frac{|A|}{R}\left(\frac{1}{2} + m\right) = \omega_0 + \frac{|A|}{R}\left(-\frac{3}{2} + n\right)} \quad (S19)$$

with degenerate eigenstates

$$|m,+\rangle = e^{im\phi}\left(e^{i\phi}p_+ + d_+\right)F(r) = e^{i(m+2)\phi}\left(e^{-i\phi}p_+ + e^{-i2\phi}d_+\right)F(r) = e^{in\phi}\left(e^{-i\phi}p_+ + e^{-i2\phi}d_+\right)F(r) = |n,+\rangle$$
$$|m,-\rangle = e^{-im\phi}\left(e^{-i\phi}p_- + d_-\right)F(r) = e^{-i(m+2)\phi}\left(e^{i\phi}p_- + e^{i2\phi}d_-\right)F(r) = e^{-in\phi}\left(e^{i\phi}p_- + e^{i2\phi}d_-\right)F(r) = |n,-\rangle$$
$$. \quad (S20)$$

Passing along the interface, the state $p_+$ increases by $2\pi$ while the term $e^{-i\phi}$ increases by $-2\pi$; i.e., there is a balance in the term $e^{-i\phi}p_+$. The term $e^{-i\phi}p_+$ can be taken as not contributing any global angular momentum to the system, which is the same case as for the terms $e^{-i\phi}p_+$, $e^{-i2\phi}d_+$, $e^{i\phi}p_-$, and $e^{i2\phi}d_-$. Thus, the phase term outside the brackets in Eq. S19, $n = m + 2$, represents the total global angular momentum. In the case of standing waves, it is best to describe the parameter $n$ as the number of nodes. The COMSOL simulation results match the theoretical analysis, as shown in **Supplementary Fig. 5**.

——**Part II: Paired Eigenstates with Broken Degeneracy**

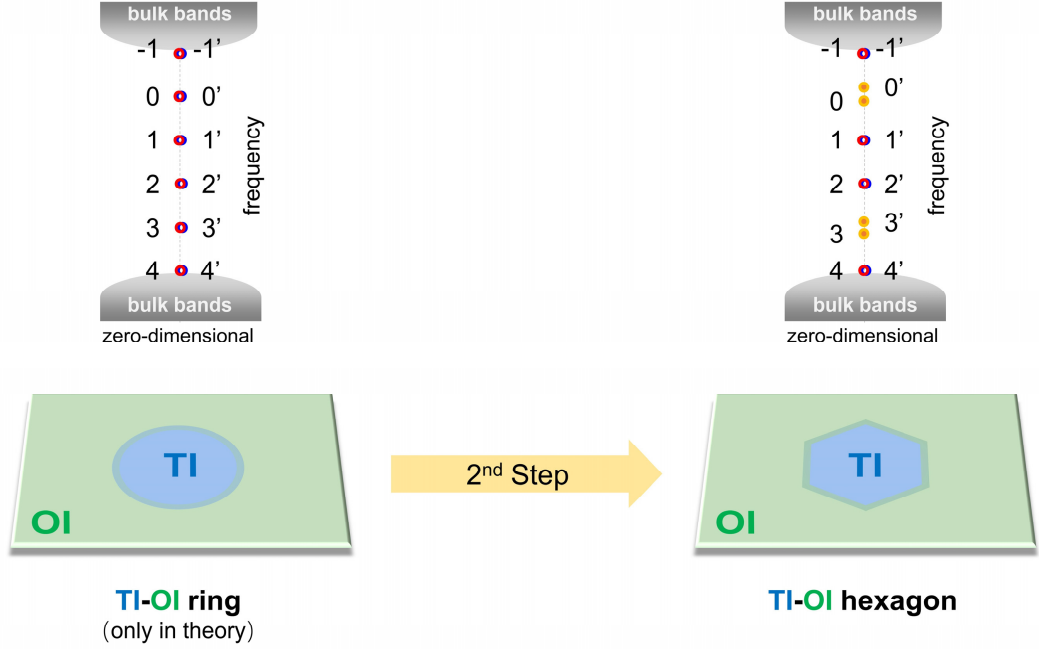

**Supplementary Fig. 3** | Paired eigenstates with broken degeneracy

We previously considered a circular geometry of the finite lattice. We now consider a hexagonal geometry of the lattice that has $C_{6v}$ symmetry. This rotational symmetry has a non-trivial effect on the eigenfunctions, as shown in Tables S1 and S2.

| symmetry operators | $\begin{pmatrix} p_+ \\ p_- \end{pmatrix}$ | $\begin{pmatrix} d_+ \\ d_- \end{pmatrix}$ | $e^{i\phi}$ |
|---|---|---|---|
| $C_6$ | $\begin{pmatrix} e^{\frac{1}{3}\pi i} & 0 \\ 0 & e^{-\frac{1}{3}\pi i} \end{pmatrix}$ | $\begin{pmatrix} e^{\frac{2}{3}\pi i} & 0 \\ 0 & e^{-\frac{2}{3}\pi i} \end{pmatrix}$ | $\phi \to \phi + \frac{1}{3}\pi$ |
| $C_3$ | $\begin{pmatrix} e^{\frac{2}{3}\pi i} & 0 \\ 0 & e^{-\frac{2}{3}\pi i} \end{pmatrix}$ | $\begin{pmatrix} e^{\frac{4}{3}\pi i} & 0 \\ 0 & e^{-\frac{4}{3}\pi i} \end{pmatrix}$ | $\phi \to \phi + \frac{2}{3}\pi$ |
| $\sigma_y$ | $\begin{pmatrix} 0 & 1 \\ 1 & 0 \end{pmatrix}$ | $\begin{pmatrix} 0 & 1 \\ 1 & 0 \end{pmatrix}$ | $\phi \to -\phi$ |
| $\sigma_x$ | $\begin{pmatrix} 0 & -1 \\ -1 & 0 \end{pmatrix}$ | $\begin{pmatrix} 0 & 1 \\ 1 & 0 \end{pmatrix}$ | $\phi \to \pi - \phi$ |

**Table S1** | Basis under $C_{6v}$ symmetry operators.

| symmetry operators | $\lvert n,+\rangle$ $= e^{in\phi}\left(e^{-i\phi}p_+ + e^{-i2\phi}d_+\right)$ | $\lvert n,-\rangle$ $= e^{-in\phi}\left(e^{i\phi}p_- + e^{i2\phi}d_-\right)$ | $\langle n,+\rvert$ $= e^{-in\phi}\left(e^{i\phi}p_- + e^{i2\phi}d_-\right)$ | $\langle n,-\rvert$ $= e^{in\phi}\left(e^{-i\phi}p_+ + e^{-i2\phi}d_+\right)$ |
|---|---|---|---|---|
| $C_6$ | $e^{in\frac{\pi}{3}}\lvert n,+\rangle$ | $e^{-in\frac{\pi}{3}}\lvert n,-\rangle$ | $e^{-in\frac{\pi}{3}}\langle n,+\rvert$ | $e^{in\frac{\pi}{3}}\langle n,-\rvert$ |
| $C_3$ | $e^{in\frac{2\pi}{3}}\lvert n,+\rangle$ | $e^{-in\frac{2\pi}{3}}\lvert n,-\rangle$ | $e^{-in\frac{2\pi}{3}}\langle n,+\rvert$ | $e^{in\frac{2\pi}{3}}\langle n,-\rvert$ |
| $\sigma_y$ | $\lvert n,-\rangle$ | $\lvert n,+\rangle$ | $\langle n,-\rvert$ | $\langle n,+\rvert$ |
| $\sigma_x$ | $e^{in\pi}\lvert n,-\rangle$ | $e^{-in\pi}\lvert n,+\rangle$ | $e^{-in\pi}\langle n,-\rvert$ | $e^{in\pi}\langle n,+\rvert$ |

**Table S2** | Eigenfunctions under C$_{6v}$ symmetry operators.

We now apply perturbation theory to study the perturbed Hamiltonian arising from the coupling of up and down (pseudo)spins, which we denote $\lvert n,+\rangle = \lvert+\rangle$ and $\lvert n,-\rangle = \lvert-\rangle$. The perturbation Hamiltonian is

$$H = \begin{pmatrix} \langle+\lvert H_s\rvert+\rangle & \langle+\lvert H_s\rvert-\rangle \\ \langle-\lvert H_s\rvert+\rangle & \langle-\lvert H_s\rvert-\rangle \end{pmatrix} = \begin{pmatrix} H_{++} & H_{+-} \\ H_{-+} & H_{--} \end{pmatrix}.$$ The matrix elements of $H$ are affected as shown in Table S3.

| symmetry operators | $H_{++}$ | $H_{--}$ | $H_{+-}$ | $H_{-+}$ |
|---|---|---|---|---|
| $C_6$ | $H_{++}$ | $H_{--}$ | $e^{-\frac{2}{3}n\pi i}H_{+-}$ | $e^{\frac{2}{3}n\pi i}H_{-+}$ |
| $C_3$ | $H_{++}$ | $H_{--}$ | $e^{-\frac{4}{3}n\pi i}H_{+-}$ | $e^{\frac{4}{3}n\pi i}H_{-+}$ |
| $\sigma_y$ | $H_{--}$ | $H_{++}$ | $H_{-+}$ | $H_{+-}$ |
| $\sigma_x$ | $H_{--}$ | $H_{++}$ | $H_{-+}$ | $H_{+-}$ |

**Table S3** | Matrix elements of $H$ under C$_{6v}$ symmetry operators.

The Hamiltonian $H$ should be unchanged under symmetry operations such that

$$\begin{aligned} H_{++} &= H_{--} \\ H_{+-} &= H_{-+} \\ H_{+-} &= e^{-\frac{2}{3}n\pi i}H_{+-} = e^{-\frac{4}{3}n\pi i}H_{+-} \end{aligned} \tag{S21}$$

We first see that the diagonal and off-diagonal terms of $H$ are the same. The off-diagonal terms vanish for the states with $n = 3k \pm 1$ (e.g., $n = 1, 2, 4, 5,...$), which means that the perturbations from the six corners balance each other and these states maintain degeneracy. However, for the states with $n = 3k$ (e.g., $n = 0, 3, 6,...$), the perturbations from the shape changes at the six corners do not balance, and the degeneracy is broken for these states.

For the states with $n = 3k$, we directly obtain the perturbed eigenfunctions:

$$\psi_1 = \frac{1}{\sqrt{2}} \begin{pmatrix} 1 & 1 \end{pmatrix} \begin{pmatrix} |+\rangle \\ |-\rangle \end{pmatrix}$$
$$\psi_2 = \frac{1}{\sqrt{2}} \begin{pmatrix} 1 & -1 \end{pmatrix} \begin{pmatrix} |+\rangle \\ |-\rangle \end{pmatrix}$$
(S22)

As an example, when $n = 0$, we have

$$\psi_1 = p_x \cos\phi + p_y \sin\phi + d_{x^2-y^2} \cos 2\phi + d_{2xy} \sin 2\phi$$
$$\psi_2 = p_y \cos\phi - p_x \sin\phi + d_{2xy} \cos 2\phi - d_{x^2-y^2} \sin 2\phi$$
(S23)

and when $n = 3$, we have

$$\psi_1 = p_x \cos 2\phi - p_y \sin 2\phi + d_{x^2-y^2} \cos\phi - d_{2xy} \sin\phi$$
$$\psi_2 = p_y \cos 2\phi + p_x \sin 2\phi + d_{2xy} \cos\phi + d_{x^2-y^2} \sin\phi$$
(S24)

All these eigenfunctions are real functions, which means that the waves are standing waves.

For other states with $n = 3k \pm 1$, the degeneracies are maintained, and the eigenfunctions have both real parts and imaginary parts, representing propagating waves.

# Supplementary Note II: Other Supplements for the Experiments

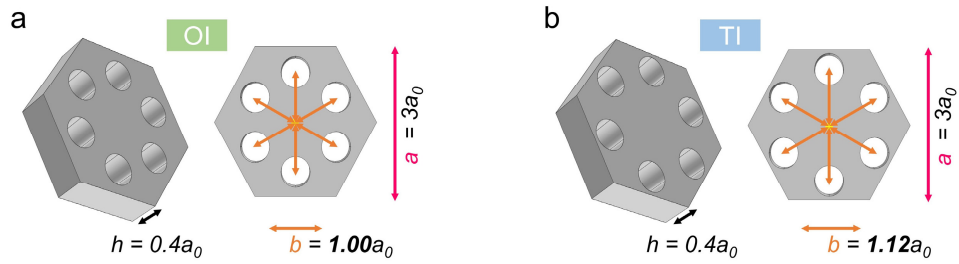

**Supplementary Fig. 4 | Structures of elastic OI and TI used in this research.**

**a:** Unit cell of an elastic OI and **b**: unit cell of a TI. The TI and OI have the same lattice constant ($a=3a_0$) in the hexagonal lattice, and each unit cell contains six of the same perforated holes in a plain plate with the same thickness. The only difference is the hole-centre distance, $b$, which equals $a_0$ (OI) and 1.12 $a_0$ (TI).

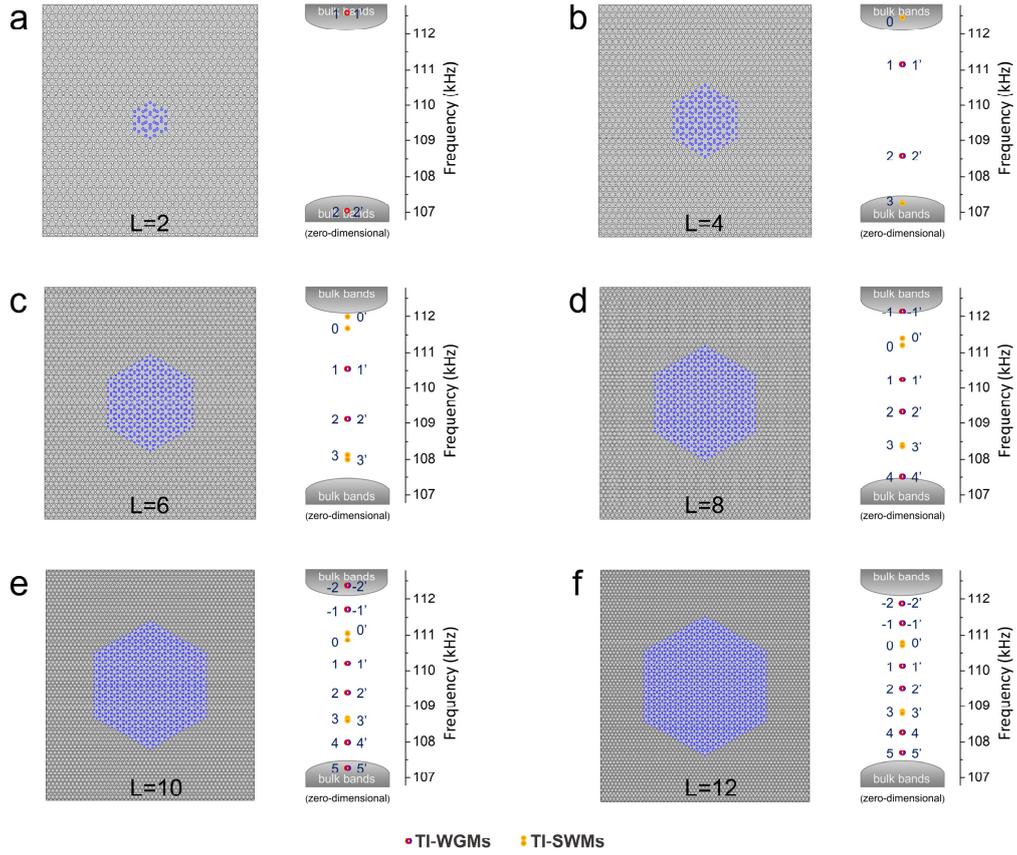

**Supplementary Fig. 5 | TI cavities of different sizes and their eigenmodes.**

For convenience of research, we set the shape of the TI particles to a regular hexagon, the size of which depends only on a single parameter, *i.e.*, the side length. When the side length is from two lattice constants to twelve lattice constants (L from 2 to 12), the cavity modes appearing in the bulk band gap of the TI/OI are as shown in **a** to **h**, respectively. No cavity mode appears in the bulk band gap when the cavity is small (*e.g.*, when L = 2). As the cavity size increases, cavity modes begin to appear in the bulk band gap, and their number continues to increase. As an example, when L = 4, cavity modes already exist in the bulk band gap, and the first modes that appear are modes 1(1') and 2(2'), whose frequencies are not split; *i.e.*, the TI-WGMs. As the cavity size further increases to a certain extent, such as when L ≥ 6, mode 0 (or 0') and mode 3 (or 3') with split frequencies (*i.e.*, the TI-SWMs) begin to appear in the bulk band gap. As the particle size further increases, an increasing number of TI-WGMs and TI-SWMs appear, and the degree of the frequency splitting of TI-SWMs decreases.

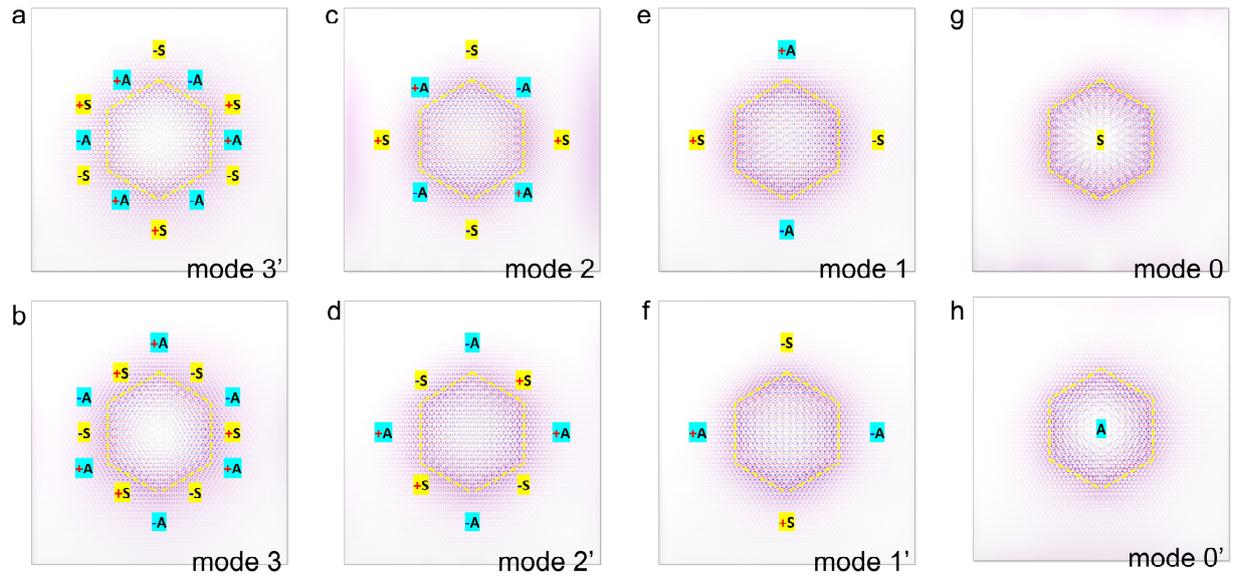

**Supplementary Fig. 6 | Paired cavity modes shown in Fig. 1 of the main text.**

**a** to **h**: Elastic field distribution of the paired modes, from low frequency to high frequency. The two modes in each pair have the same spatial rotational symmetry about the centre of the hexagon cavity. Therefore, naming these paired cavity modes according to their symmetry index, from low frequency to high frequency, as modes 3(3'), 2(2'), 1(1'), and 0(0') is advantageous.

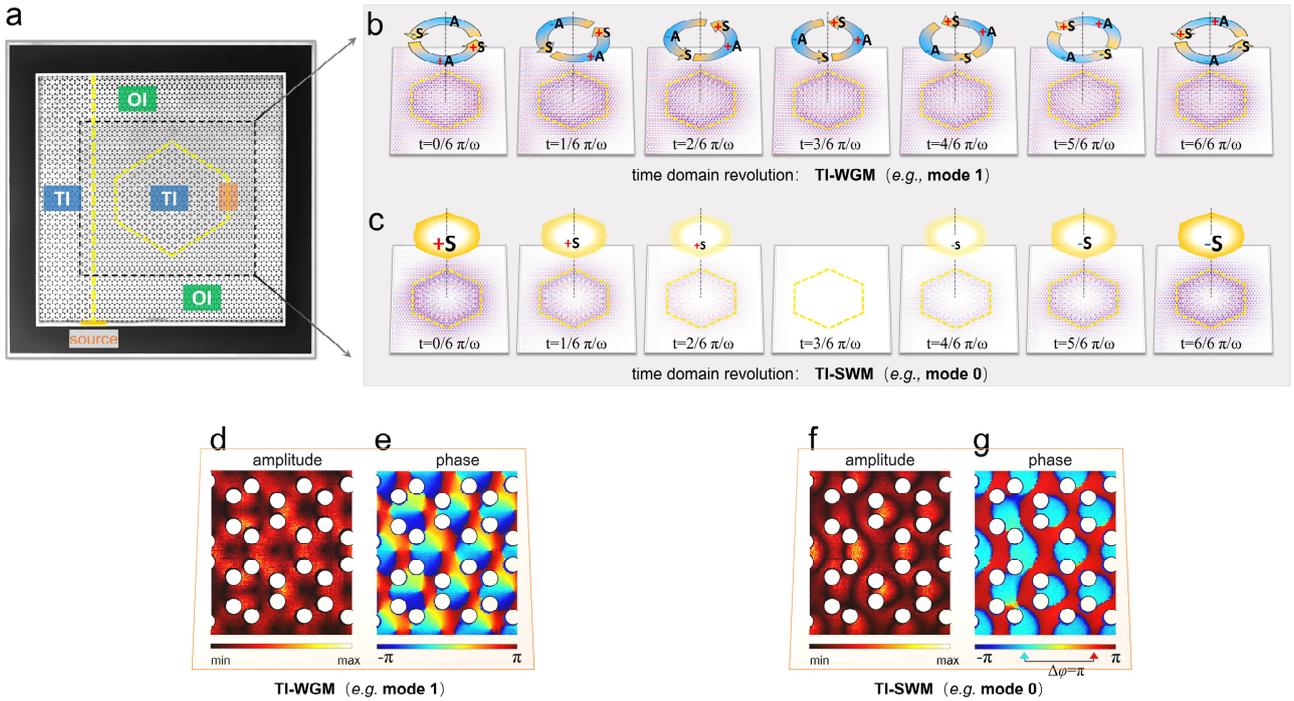

**Supplementary Fig. 7 | Cavity mode excitations of two different categories.**

**a**: Sample image and experimental configuration, similar to the cases shown in **Figs. 3** and **4** of the main text. **b**: Calculated elastic displacement field distribution of the TI cavity in the half resonance period when a TI-WGM (*e.g.*, mode 1 (1')), is excited. Elastic energy is mainly concentrated on the boundary of the TI cavity and exists in the form of a travelling wave. The elastic wave on the cavity boundary is similar to that in the case of a helical edge state with spin $+½$ in the straight waveguide, evolving in the form of "$+S$ — $+A$ — $-S$ — $-A$" in the time domain. $S$ and $A$ are the base vectors of the TI-OI helical edge states (*i.e.*, $S \pm iA$ with spin $\pm½$). **c**: Calculated elastic displacement field distribution of the TI cavity in the half resonance period when a TI-SWM (*e.g.*, mode 0) is excited. Although the elastic energy is still mainly concentrated on the boundary of the TI cavity, the elastic waves collectively oscillate throughout the cavity, including at the boundaries. **d** and **e**: Experimentally measured field distributions of the amplitude and phase, respectively, of the elastic waves inside the TI cavity (orange box shown in **a**) when mode 1 is excited. The phase changes continuously, following the characteristics of travelling waves. **f** and **g**: Experimentally measured field distributions of the amplitude and phase, respectively, when mode 0 is excited. Here, the phase change is not continuous but has an island-like distribution with a phase difference of 180°, which is a standard feature of standing waves.

39 / 39